# The reducing role of hydrogen peroxide on the formation of gold nanostructures in aqueous microdroplets with dissolved tetrachloroaurate ions

Zhaoyuan Liu, Renze Yu, Xin Wang, Qiang Chen[*]

Institute of Electromagnetics and Acoustics, Fujian Provincial Key Laboratory of Plasma and Magnetic Resonance, Key Laboratory of Electromagnetic Wave Science and Detection Technology, Xiamen University, Xiamen 361005, China

[*]Email: chenqiang@xmu.edu.cn

A recent article by Lee et al.[1] in Nature Communications has reported an intriguing phenomenon that gold nanostructures (AuNSs) can be spontaneously formed in aqueous microdroplets with dissolved tetrachloroaurate ions ($AuCl_4^-$). The authors suggested three possible electron donors for the reduction of $AuCl_4^-$, including the strong electric field at the microdroplet's surface, the hydroxyl ions, and the $AuCl_4^-$ itself ($AuCl_4^- \rightarrow AuCl_2^- + Cl_2$). However, we find that the hydrogen peroxide ($H_2O_2$) spontaneously produced at the microdroplets[2-4] might be also responsible for the $AuCl_4^-$ reduction.

The microdroplets in the article of Lee et al.[1] were generated by a high-pressure, dry $N_2$ at 120 psi. The AuNSs were observed in the collection of a fusion of aqueous $AuCl_4^-$ microdroplets and the water microdroplets, and it was also found in the collection of aqueous $AuCl_4^-$ microdroplets. To form AuNSs from the reduction of $AuCl_4^-$, electrons should be provided somehow. Three possible electron sources were suggested in Ref. [1] as mentioned above. We noted that in a recent paper of Lee et al.[5], spontaneous reduction of several organic molecules in aqueous microdroplets was also observed, and they pointed out that the charge separation of $OH^-$ ($OH^- \rightarrow OH + e$) is likely the electron source for the reduction. The charge separation has been suggested to be induced by a strong intrinsic electric field spontaneously induced at the microdroplet's surface[6]. Due to the high rate constant of OH+OH→$H_2O_2$ ($k$=5.5×10$^9$ M$^{-1}$ s$^{-1}$)[7], this charge separation certainly leads to a spontaneous formation of $H_2O_2$ in the micro-sized water droplets, which has been reported by Lee et al.[4]. In the case of organic molecules, it might be the charge separation which provides the electrons, while we found that in the $AuCl_4^-$ reduction, another electron donor is not negligible, i.e., the $H_2O_2$ spontaneously formed at the microdroplets[2-4].

Previously, we have investigated the formation of gold nanoparticles using a discharge plasma to irradiate an aqueous solution of $AuCl_4^-$. The $H_2O_2$ generated by the plasma-liquid interactions[8-11] was confirmed to be an important reductant for reducing $AuCl_4^-$ to Au (0). We have proposed Eqs. 1-3 for the $AuCl_4^-$ reduction in an aqueous solution containing $H_2O_2$ and OH radicals. A simple calculation indicated that the standard Gibbs free energy change per mole of reaction are -86.85 kJ/mol, -332.25 kJ/mol, and -11.58 kJ/mol for Eqs. 1-3, respectively. The negative values of the Gibbs free energy mean that these reactions can be proceeded spontaneously[12].

$$2AuCl_4^- + 3H_2O_2 \rightarrow 2Au + 3O_2 + 6H^+ + 8Cl^- \tag{1}$$

$$2AuCl_4^- + 3H_2O_2 + 6OH^- \rightarrow 2Au + 6H_2O + 6O_2 + 8Cl^- \tag{2}$$

$$2AuCl_4^- + 6H_2O + 3OH \rightarrow 2Au + 3H_2O_2 + 6H^+ + 8Cl^- \tag{3}$$

Interestingly, Lee et al.[1] also observed oxygen evolution during the AuNSs formation as indicated in Eqs. 1-2. The existence of oxygen evolution implies that reaction 3 should be ruled out from the possible reduction reactions of $AuCl_4^-$.

To confirm the AuNSs formation of the $AuCl_4^-$ reduction by $H_2O_2$, we performed experiment by mixing aqueous solution of $AuCl_4^-$ and $H_2O_2$ in a cuvette. Figure 1 shows the temporal photos of the aqueous solution of $AuCl_4^-$/PVP with and without $H_2O_2$ [Polyvinylpyrrolidone (PVP) was used to stabilize the formed nanoparticles]. Obviously, the transparent aqueous solution of $AuCl_4^-$/PVP with $H_2O_2$ gradually changed to apricot (See supplementary Materials for the experimental detail and video for the change of the mixed solution). Gold nanoparticles formation is indicted by the existence of the gold surface plasmon resonance peak at 539 nm in the absorbance of the mixed solution 20 min after the mixture (Fig. 2). Measurements of scanning electron microscopy (SEM) and energy dispersive X-ray spectroscopy (EDS) further confirmed the formed particles are gold nanoparticles (Fig. 3).

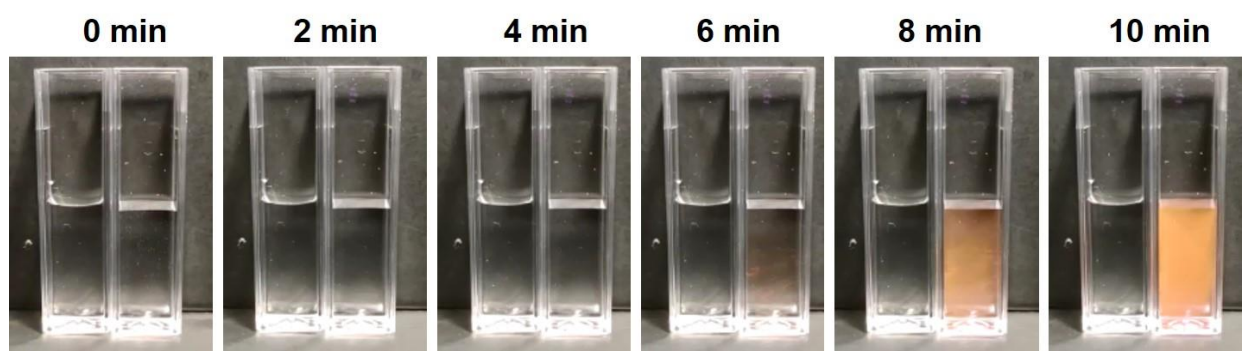

**Fig. 1 Temporal photos of the mixed solution of $AuCl_4^-$/PVP and $H_2O_2$.** Photos were screenshotted from a video taken after the mixing of the solution (See Supplementary Information).

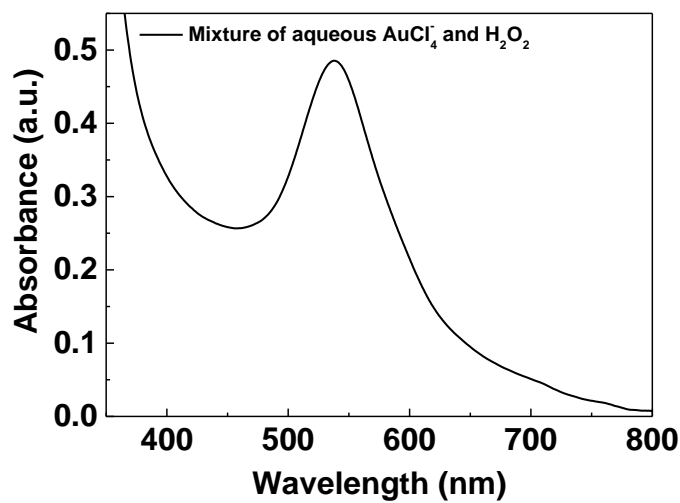

**Fig. 2 Absorbance of the mixed solution of $AuCl_4^-$/PVP and $H_2O_2$ after 20 min of the mixing**

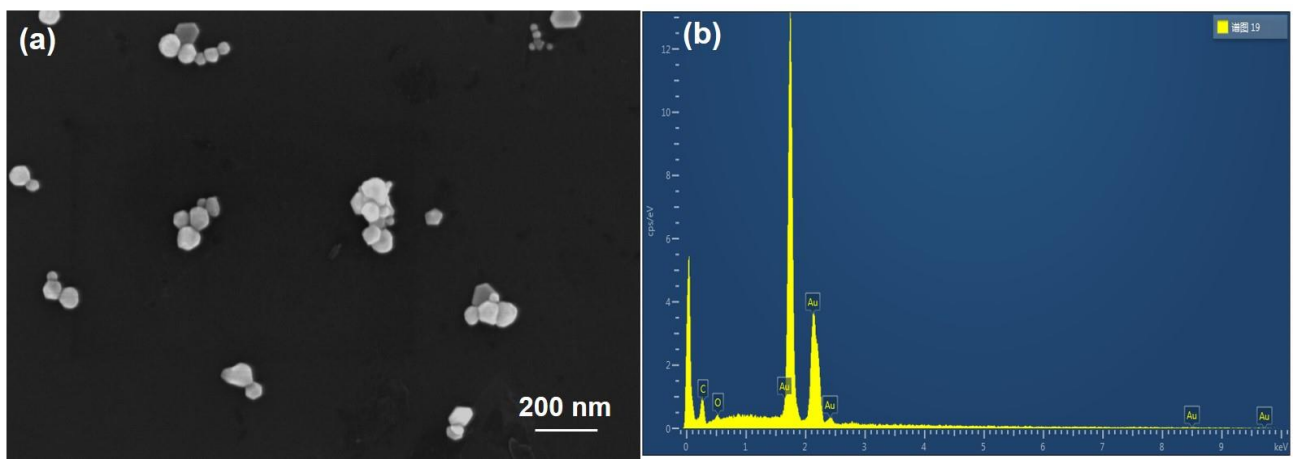

**Fig. 3 SEM and EDS results of nanoparticles formed by mixing $AuCl_4^-$/PVP and $H_2O_2$.** The mixture was made by directly dropping 1 ml $H_2O_2$ (30%, w/w) into 1 ml aqueous $AuCl_4^-$/PVP solution, and the concentrations of $AuCl_4^-$ and PVP were 0.25 mM, respectively.

Finally, we conclude that the $H_2O_2$ spontaneously produced at or near the surface of aqueous microdroplets plays an important role for reducing $AuCl_4^-$ to form gold nanostructures in the microdroplets with $AuCl_4^-$.

**Data availability**

Supplementary Information: the experimental details and the video of the reaction for aqueous $AuCl_4^-$ with $H_2O_2$.

**Acknowledgements**

This work was supported by National Natural Science Foundation of China (Grant No.: 52077185).

Q.C thanks Dr. Kai Luo in Xiamen University for his insightful discussions.

**Author information**

These authors contributed equally: Zhaoyuan Liu, Renze Yu.

Affiliations

Institute of Electromagnetics and Acoustics, Fujian Provincial Key Laboratory of Plasma and Magnetic Resonance, Key Laboratory of Electromagnetic Wave Science and Detection Technology, Xiamen University, Xiamen 361005, China

Contributions

Z.L. and R.Y. performed the experiments. X.W. performed the SEM and EDS measurement. Q.C wrote the manuscript and is responsible for the conception. All authors approved the submission of the manuscript.

**Ethics declarations**

Competing interests

The author declares no competing interests.

# Supplementary Information

# The reducing role of hydrogen peroxide on the formation of gold nanostructures in aqueous microdroplets with dissolved tetrachloroaurate ions

Zhaoyuan Liu, Renze Yu, Xin Wang, Qiang Chen[*]
Institute of Electromagnetics and Acoustics, Fujian Provincial Key Laboratory of Plasma and Magnetic Resonance, Key Laboratory of Electromagnetic Wave Science and Detection Technology, Xiamen University, Xiamen 361005, China
[*]Email: chenqiang@xmu.edu.cn

**Experimental details**

Hydrogen peroxide ($H_2O_2$, 30%, w/w) was purchased from Xilong Scientific Co., Ltd. Polyvinylpyrrolidone (PVP) [$(C_6H_9NO)_n$, average molecular weight 58000, K29-32], Hydrogen tetrachloroaurate (III) trihydrate ($HAuCl_4 \cdot 3H_2O$), ACS, 99.99%, was purchased from Alfa Aesar.

For the absorbance measurement, 1 ml aqueous solution of $HAuCl_4$/PVP (0.25 mM/0.25 mM) was prepared in a quartz cell and then 1 ml $H_2O_2$ (30%, w/w) was dropped into the cell. Absorbance of the mixed aqueous solution was detected in the quartz cell using by Ocean Optics spectrometer (USB2000+) with a light source of HL-2000 (360 nm-2000 nm).

A video was taken for comparison of the aqueous solution of $HAuCl_4$/PVP (0.25 mM/0.25 mM) and a mixture of $HAuCl_4$/PVP (0.25 mM/0.25 mM) and $H_2O_2$ (30%, w/w). One can find that the solution without $H_2O_2$ was transparent, while the solution with $H_2O_2$ turned to apricot, which implies the formation of gold nanoparticles.